\documentclass[a4paper,11pt]{article}
\usepackage{amsmath,graphicx,color,mathrsfs,subfigure,multicol,hyperref,aas_macros,xspace}

\newcommand*{\ie} {\emph{i.\,\!e.}\xspace}

\newcommand{\lone}{\ensuremath{{\ell_1}}\xspace}
\newcommand{\ltwo}{\ensuremath{{\ell_2}}\xspace}
\newcommand{\ltre}{\ensuremath{{\ell_3}}\xspace}
\newcommand{\ldep}{\ensuremath{{\lone\ltwo\ltre}}\xspace}
\newcommand*{\lmax} {\ensuremath{\ell_\text{max}}\xspace}
\newcommand*{\avg} [1] {\ensuremath{\left\langle\,{#1}\,\right\rangle}\xspace}
\newcommand{\threej}[6]{%
\mbox{$\begin{pmatrix}{#1\!}&{#2\!}&{#3}\\{#4\!}&{#5\!}&{#6}\end{pmatrix}$}}

\def\circa#1{\,\raise.3ex\hbox{$#1$\kern-.75em\lower1ex\hbox{$\sim$}}\,}
\makeatletter

\def\e{M^2}
\def\M{M_P}
\def\g{\gamma}
\def\matter{{\rm brane}}
\def\p{\phi}

\def\fnleq{f_{\rm NL}^{\rm eq}}

\def\fnlorth{f_{\rm NL}^{\rm orth}}
\def\fnlgrad{f_{\rm NL}^{\rm grad}}
\def\fnltime{f_{\rm NL}^{\rm time}}

\def\hfnleq{\hat{f}_{\rm NL}^{\rm eq}}
\def\hfnlorth{\hat{f}_{\rm NL}^{\rm orth}}

\def\nn{\nonumber}

\def\k{{\bf k}}

\def\g{\gamma}
\def\c{c_D}

\def\e{M^2}
\def\M{M_P}

\def\sd{\dot \sigma}

\begin{document}

\date{\mbox{}}

\title{
\vspace{-2.0cm}
\vspace{2.0cm}
{\bf \huge  Orthogonal non-Gaussianity in DBI Galileon: \\
\LARGE prospect for Planck polarisation and post-Planck experiments}
 \\[8mm]
}

\author{
Kazuya Koyama$^1$, Guido Walter Pettinari$^{1,2}$, Shuntaro Mizuno$^{3,4}$, Christian Fidler$^{1}$
\\[8mm]
\normalsize\it
$^1$ Institute of Cosmology \& Gravitation, University of Portsmouth,\\
\normalsize\it Dennis Sciama Building, Portsmouth, PO1 3FX, United
Kingdom \\
$^2$
\normalsize\it
Department of Physics \& Astronomy, University of Sussex,\\
\normalsize\it
Brighton, BN1 9QH,
United Kingdom \\
$^3$
\normalsize\it
APC (CNRS-Universit\'e Paris 7),
10 rue Alice Domon et L\'eonie Duquet, \\
\normalsize\it
75205 Paris Cedex 13,
France \\
$^4$
\normalsize\it
Laboratoire de Physique Th\'eorique,
Universit\'e Paris-Sud 11 et CNRS, \\
\normalsize\it
B\^atiment 210, 91405 Orsay
\normalsize\it
Cedex, France \\
}
\maketitle

\setcounter{page}{1}
\thispagestyle{empty}

\begin{abstract}
\noindent
In this note, we study cosmic microwave background (CMB) constraints on primordial non-Gaussianity in DBI galileon models in which an induced gravity term is added to the Dirac-Born-Infeld (DBI) action. In this model, the non-Gaussianity of orthogonal shape can be generated. We provide a relation between theoretical parameters and orthogonal/equilateral non-linear parameters using the Fisher matrix approach for the CMB bispectrum. In doing so, we include the effect of the CMB transfer functions and experimental noise properties by employing the recently developed \textsf{SONG} code. The relation is also shown in the language of effective theory so that it can be applied to general single-field models.
Using the bispectrum Fisher matrix and the central values for equilateral and orthogonal non-Gaussianities found by the Planck temperature survey, we provide forecasts on the theoretical parameters of the DBI galileon model. We consider the upcoming Planck polarisation data and the proposed post-Planck experiments COrE and PRISM.
We find that Planck polarisation measurements may provide a hint for a non-canonical sound speed at the 68$\%$ confidence level. COrE and PRISM will not only confirm a non-canonical sound speed but also exclude the conventional DBI inflation model at more than the 95$\%$ and 99$\%$ confidence level respectively, assuming that the central values will not change. This indicates that improving constraints on non-Gaussianity further by future CMB experiments is invaluable to constrain the physics of the early universe.
\end{abstract}

\section{Introduction}
Primordial non-Gaussianity of the curvature perturbation provides
valuable information on the physics in the very early Universe
\cite{Komatsu:2010hc}. Non-linearity of quantum fluctuations during
inflation gives rise to a bispectrum in the Cosmic Microwave Background
(CMB) temperature anisotropies that peaks for equal-side triangles
\cite{Koyama:2010xj}. The most popular model for the equilateral type
non-Gaussianity is the Dirac-Born-Infeld (DBI) inflation model
\cite{Alishahiha:2004eh}. In the DBI inflation model, the inflaton is
identified as the position of a D3-brane in a higher dimensional
spacetime. The DBI action that describes the motion of the brane is a
non-linear function of the kinetic term of the inflaton, which leads to
the non-Gaussianity of quantum fluctuations. While the original DBI inflation model
considered only the motion and fluctuations in the radial direction, we can consistently
take into account the dynamics and fluctuations in the angular directions \cite{Langlois:2008wt}
(see also \cite{Langlois:2008qf,Arroja:2008yy}). Although the original single field DBI inflation
model \cite{Alishahiha:2004eh} is under strain from an additional requirement related with
the compactification scheme of the string theory \cite{Baumann:2006cd,Lidsey:2007gq},
this can be evaded in multi-field DBI inflation models and the shape of the bispectrum remains
as in the single-field model \cite{Langlois:2008wt}.

Recently, a natural extension of the DBI inflation model has been
obtained by adding an induced gravity term \cite{RenauxPetel:2011dv, RenauxPetel:2011uk}. This leads to the quartic galileon Lagrangian \cite{Nicolis:2008in} when the motion of the brane is non-relativistic. Thus this model is dubbed as the DBI galileon model \cite{DBIg}. This is one of a very few models where the non-Gaussianity of orthogonal shape can be generated \cite{RenauxPetel:2011dv}. The orthogonal shape of non-Gaussianity was originally discovered in the context of the effective theory of inflation \cite{Cheung:2007st}, which has a minimum overlap between local and equilateral non-Gaussianities \cite{Senatore:2009gt}. In the WMAP nine-year (WMAP9) data, a hint was found that the orthogonal type non-Gaussianity could be non-zero at the 2$\sigma$ level when the equilateral non-Gaussianity is included in the parameter space \cite{Bennett:2012fp}. On the other hand, the Planck satellite found no evidence of any type of non-Gaussianity \cite{planck-ng}.

In this note, we provide a relation between the equilateral and orthogonal templates, parametrised respectively by the non-linear parameters $\fnleq$ and $\fnlorth$, and the theoretical parameters in the DBI galileon models by properly taking into account the CMB transfer functions and experimental noise properties. We then derive the constraints on $\fnleq, \fnlorth$ from Planck and provide forecasts for Planck polarisation and post-Planck experiments, Cosmic Origins Explore (COrE) \cite{core-collaboration:2011} and Polarized Radiation Imaging and Spectroscopy Mission (PRISM) \cite{prism-collaboration:2013b}, including the possibility of the simultaneous presence of the two shapes in the data. We then provide forecasts for constraints on the parameters in the DBI galileon model. We also present these forecasts in the effective theory language so that they can be easily applied to more general single-field models.

This paper is organised as follows. In section 2, we summarise the prediction of non-Gaussianity in the DBI galileon models. In section 3, we present the equilateral and orthogonal templates and discuss the overlap between theoretical bispectrum shapes and these templates. In section 4, we apply the Planck results to obtain constraints on the theoretical parameters and provide forecasts for constraints from Planck polarisation.
We study forecasts for the post-Planck experiments COrE and PRISM in section 5. In particular, we provide a relation between templates and theoretical parameters using the exact Fisher matrix, which is specific to COrE and PRISM. Using the bispectrum Fisher matrix, we provide forecasts on parameters in the DBI galileon model and in the effective theory. Section 6 is devoted to the conclusion.

\section{Non-Gaussianity in DBI galileon model}
In this section, we summarise the set-up of the DBI galileon model and its predictions for the non-Gaussianity by following Refs.~\cite{RenauxPetel:2011dv, RenauxPetel:2011uk}.
We consider a D3-brane with tension $T_3$ evolving in a 10-dimensional geometry described by the metric
\begin{equation}
ds^2 = h^{-1/2}(y^K)\,g_{\mu \nu}dx^\mu dx^\nu + h^{1/2}(y^K)\, G_{IJ}(y^K)\, dy^I dy^J \equiv H_{AB} dY^A dY^B,
\end{equation}
with coordinates $Y^A=\left\{x^\mu, y^I\right\}$,
where $\mu=0,\ldots 3$ and $I=1,\ldots, 6$. The induced metric on the 3-brane is given by
\begin{equation}
\gamma_{\mu \nu
} =H_{AB} \partial_\mu Y_{\rm (b)}^A \partial_\nu Y_{\rm (b)}^B,
\end{equation}
where the  brane embedding is defined by the functions  $Y_{\rm (b)}^A(x^\mu)$, with $x^\mu$ being the spacetime coordinates on the brane. We choose the brane embedding as $Y_{\rm (b)}^A = (x^\mu, \varphi^I(x^{\mu}))$. Then the induced metric can be written as
\begin{equation}
\gamma_{\mu \nu} = h^{-1/2} \left( g_{\mu \nu}  + h \, G_{IJ} \partial_\mu \varphi^I \partial_\nu \varphi^J \right)
\label{induced-metric}\,.
\end{equation}
The action in the DBI galileon model is given by
\begin{equation}
S= \int {\rm d}^4 x \left[ \frac{\M^2}{2}  \sqrt{-g} R[g]  +\frac{\e}{2} \sqrt{-\g} R[\g] + \sqrt{-g}  {\cal L}_\matter    \right],
\label{action}
\end{equation}
where $\M$ and $M$ are constant mass scales and
\begin{equation}
 {\cal L}_\matter= -\frac{1}{f(\phi^I)}\left(\sqrt{{\cal D}}-1\right) -V(\phi^I)\,.
 \label{brane-action}
\end{equation}
The second term in the action is the induced gravity term, which is absent in the conventional DBI inflation model.
Here, we have introduced the rescaled variables using the tension of the D3 brane, $T_3$,
\begin{equation}
f= \frac{h}{T_3} \; , \qquad \phi^I = \sqrt{T_3}\varphi^I\,,
\label{redef}
\end{equation}
we included potential terms in the brane action and we defined
\begin{equation}
{\cal D} \equiv  \det(\delta^{\mu}_{\nu}+f \, G_{IJ} g^{\mu \rho}  \partial_{\rho} \p^I \partial_{\nu} \p^J )\,,
\label{Ddef}
\end{equation}
where $G_{IJ}(\phi^K)$ will play the role of a metric in the space of the scalar fields $\phi^I$. By defining the mixed kinetic terms for the scalar fields
\begin{equation}
X^{IJ} \equiv -\frac12 g^{\mu \nu}\partial_{\mu} \phi^I \partial_{\mu} \phi^J\,,
\label{def-XIJ}
\end{equation}
it has been shown that the explicit expression of ${\cal D}$ reads \cite{Langlois:2008wt}
\begin{equation}
{\cal D}=1-2f G_{IJ}X^{IJ}+4f^2 X^{[I}_IX_J^{J]} -8f^3 X^{[I}_IX_J^{J} X_K^{K]}+16f^4 X^{[I}_IX_J^{J} X_K^{K}X_L^{L]},
\label{def_explicit}
\end{equation}
where the brackets denote antisymmetrisation on the field indices and $X_{I}^{J}=G_{IK} X^{KJ}$. Similarly, one can express $ \sqrt{-\g}R[\g]$ in terms of the fields and the geometrical quantities associated to the cosmological metric, leading to a multifield relativistic extension of the quartic galileon Lagrangian in curved spacetime.

In this paper, for simplicity, we only consider the single-field model where we can ignore the dynamics of angular directions and the late time curvature perturbation is dominated by the radial fluctuations. There are two parameters in the single field model. One is the background value of $\cal D$:
\begin{equation}
\c^2 \equiv 1- f  \dot \sigma^2\,,
\end{equation}
where $\sd \equiv \sqrt{G_{IJ} \dot \phi^I   \dot \phi^J}$ plays the role of an effective collective velocity of the fields. In the DBI inflation model, $c_D$ corresponds to the sound speed of the perturbations, $c_s$. The other parameter characterises the effect of the induced gravity
\begin{equation}
\alpha \equiv \frac{ f H^2 M^2}{\c^2 \sqrt{h}}\,,
\label{def-alpha}
\end{equation}
where $H$ is the Hubble parameter. If $\alpha=0$ we reproduce the DBI inflation model. The parameter $\alpha$ is restricted to be $0 \leq \alpha \leq 1/9$ for $c_D \ll 1$ to ensure that the fluctuations are not ghosts. 

We only show the final results for the non-Gaussianity of the gravitational potential. Detailed calculations can be found in Refs.~\cite{RenauxPetel:2011dv, RenauxPetel:2011uk}.
The bispectrum of the Newtonian potential $\Phi$ has the form
\begin{equation}
  \langle \Phi_{\vec k_1} \Phi_{\vec k_2} \Phi_{\vec k_3} \rangle
  = (2\pi)^3 \, \delta^{(3)}(\vec{k}_1+\vec{k}_2+\vec{k}_3) \,\, S(k_1,k_2,k_3)\ \;,
  \label{phibispectrum}
\end{equation}
where $S$ is the primordial bispectrum shape and the Dirac delta enforces spatial homogeneity.
In the single-field DBI galileon model, two bispectrum shapes arise \cite{Senatore:2009gt}:
\begin{eqnarray}\label{eq:shapeone}
&&S^{\rm (grad)}(k_1,k_2,k_3)=-\frac{27}{17}\; \fnlgrad \Delta_\Phi^2 \\ \nonumber
&&\quad\qquad\times\frac{\left(
24 K_3{}^6- 8 K_2{}^2 K_3{}^3 K_1- 8 K_2{}^4 K_1{}^2+22 K_3{}^3 K_1{}^3- 6 K_2{}^2 K_1{}^4+ 2 K_1{}^6\right)}{K_3{}^9K_1{}^3}\ , \\ \nonumber
&&S^{\rm (time)}(k_1,k_2,k_3)=162\;
\fnltime \Delta_\Phi^2\cdot\frac{1}{K_3{}^3K_1{}^3}\ .
\end{eqnarray}
where
\begin{eqnarray}
K_1 &=& k_1 + k_2 + k_3\ , \\ \nn
K_2 &=& \left( k_1 k_2 + k_2 k_3 + k_3 k_1 \right)^{1/2}\ , \\ \nn
K_3 &=& \left( k_1 k_2 k_3 \right)^{1/3}\ ,
\end{eqnarray}
and we used the definition of the power spectrum
\begin{equation}
\langle\Phi(\vec k_1)\Phi(\vec k_2)\rangle=(2\pi)^3\delta^{(3)}(\vec k_1+\vec k_2)\frac{\Delta_\Phi}{k^3}\ .
\end{equation}
These two shapes arise from the two distinct cubic interactions of the comoving curvature perturbation $\zeta$; the first shape $S^{\rm (grad)}$ arises from $\dot{\zeta} (\partial_i \zeta)^2$ while the second shape $S^{\rm (time)}$ arises from $\dot{\zeta}^3$ \cite{Senatore:2009gt}. The amplitudes of these bispectrum are determined by
$\c$ and $\alpha$ as
\begin{eqnarray}
\fnlgrad &=& \frac{85}{324} A^{\rm grad}
\left(1 -\frac{1}{c_D^2} \right), \quad A^{\rm grad} = \lambda^2 \;\; \frac{1 - \alpha(9 - 2 c_D^2 - 3 \lambda^2 )}{1- 3 \alpha (3 - 2 c_D^2)}, \nonumber\\
\fnltime &=&
\frac{5}{81} A^{\rm time} \left(1 -\frac{1}{c_D^2} \right), \quad
A^{\rm time} = \frac{1- 3 \alpha (5 - 2 c_D^2 - 4 \lambda^2 + \lambda^4)}{
1- 3 \alpha (3 - 2 c_D^2)},
\label{DBIg}
\end{eqnarray}
where $\lambda$ is the ratio between the angular
and radial speed of sound
\begin{equation}
\lambda = \sqrt{\frac{1 - 3 \alpha ( 3 - 2 c_D^2)}{1 - \alpha(5 - 2 c_D^2)}}.
\end{equation}
For $\alpha=0$, we recover multi-field DBI inflation models in which the radial and angular sound speeds are the same, $\lambda=1$.  

\section{CMB temperature and polarisation bispectrum and templates}
\label{sec:cmb_bispectra}

The shapes of the bispectrum described in section 2 are not factorisable thus it is numerically challenging to construct optimal estimators. Instead, the WMAP collaboration has been using the following equilateral \cite{Creminelli:2005hu} and orthogonal \cite{Senatore:2009gt} templates
\begin{eqnarray}
\label{eq:equilateral_orthogonal_templates}
S^{\rm (eq)}(k_1,k_2,k_3) &=& \fnleq \cdot6\Delta_{\Phi}^2\cdot\left(-\frac{1}{k_1^3k_2^3}-\frac{1}{k_1^3k_3^3}-\frac{1}{k_2^3k_3^3}-\frac{2}{k_1^2k_2^2k_3^2}+\frac{1}{k_1 k_2^2k_3^3}+(5\ perm.)\right), \\\notag
S^{\rm (orth)}(k_1,k_2,k_3) &=&\fnlorth \cdot6\Delta_{\Phi}^2\cdot\left(-\frac{3}{k_1^3k_2^3}-\frac{3}{k_1^3k_3^3}-\frac{3}{k_2^3k_3^3}-\frac{8}{k_1^2k_2^2k_3^2}+\frac{3}{k_1 k_2^2k_3^3}+(5\ perm.)\right),
\end{eqnarray}
and gave constraints on the non-linear parameters $(\fnleq, \fnlorth)$.  In order to constrain the amplitude of the galileon shapes in Eq.~\eqref{eq:shapeone}, we first need to relate $(\fnlgrad, \fnltime)$ to the observed $(\fnleq, \fnlorth)$.

To this purpose, we calculate the bispectrum of the CMB temperature and polarisation anisotropies for all the considered shapes. In harmonic space, this is defined as the three-point function
\begin{equation}
  \langle a^X_{\ell_1 m_1}\; a^Y_{\ell_2 m_2}\; a^Z_{\ell_3 m_3} \rangle \;,
\end{equation}
where the field indices $X,Y,Z$ denote either temperature ($T$) or E-mode polarisation ($E$).
The $a^X_{\ell m}$ are the coefficients of the spherical harmonics decomposition of the CMB map for the $X$ field. Given a cosmological model, their value can be predicted as
\begin{equation}
  a^X_{\ell m}= 4 \pi (-i)^{\ell} \int \frac{d^3 \k}{(2 \pi)^3}
  \Phi( \k) \Delta^X_{\ell}(k) Y^{*}_{\ell m}(\hat{\k}) \;,
\end{equation}
where $\Delta^X_{\ell}$ is the radiative transfer function for the field $X$, which is obtained by solving the Boltzmann-Einstein system of differential equations at first order \cite{ma:1995a, seljak:1996a}.
The statistical isotropy of the Universe allows us to parametrise the directional dependence of the three-point function via the Wigner 3$j$-symbol \cite{komatsu:2002a}, and thus define the angle-averaged bispectrum $\,B^{XYZ}_{\ell_1 \ell_2 \ell_3}\,$ as
\begin{equation}
\langle a^X_{\ell_1 m_1}\; a^Y_{\ell_2 m_2}\; a^Z_{\ell_3 m_3} \rangle
  = \threej{\lone}{\ltwo}{\ltre}{m_1}{m_2}{m_3} \; B^{XYZ}_{\ell_1 \ell_2 \ell_3}.
\end{equation}
We compute the angle-averaged bispectrum by projecting the primordial bispectrum for the Newtonian potential in Eq.~(\ref{phibispectrum}) on the sky today \cite{Komatsu:2001a, Fergusson:2007a}:
\begin{eqnarray}
	\label{eq:bispectrum_integral}
	B^{XYZ}_\ldep \, = h_\ldep\, \left(\frac{2}{\,\pi}\right)^3 \int dr\, r^2\,
  \int dk_1\,dk_2\,dk_3\,
  \left(k_1\,k_2\,k_3\right)^2\, S(k_1,k_2,k_3)\, \\[0.15cm]
  \nonumber
  \,j_{\ell_1}(r k_1)\,\Delta^X_{\ell_1}(k_1)
  \,j_{\ell_2}(r k_2)\,\Delta^Y_{\ell_2}(k_2)
  \,j_{\ell_3}(r k_3)\,\Delta^Z_{\ell_3}(k_3) \;,
\end{eqnarray}
where $j_\ell$ is the spherical Bessel function of order $\ell$ and
\begin{align*}
  h_\ldep \,=\, \sqrt{\frac{(2\lone+1)(2\ltwo+1)(2\ltre+1)}{4\pi}}
  \,\threej{\lone}{\ltwo}{\ltre}{0}{0}{0} \;.
\end{align*}
is the purely geometrical factor that forces $\lone+\ltwo+\ltre$ to be even, as expected from the even-parity fields $T$ and $E$.


Following Refs.~\cite{babich:2004a, yadav:2008a, Komatsu:2001a}, we define the 2D scalar product of two angular bispectra, or Fisher matrix element, as
\begin{equation}
  B^{(i)} \cdot B^{(j)} =
    \sum\limits_{ABC,XYZ}\;\sum\limits_{\ell_1\ell_2\ell_3}^{\ell_{\rm max}}
    B^{(i),ABC}_\ldep \; \left(\text{\bf Cov}^{-1}\right)^{ABC,XYZ}_\ldep\; B^{(j),XYZ}_\ldep \;.
\end{equation}
The first sum involves all possible pairs of the eight bispectra (TTT, TTE, TET, ETT, EET, ETE, TEE, EEE), for a total of 64 addends. The latin indices refer to the four types of bispectra considered in this paper: $(i),(j)=\text{eq},\,\text{orth},\,\text{grad},\,\text{time}$.
The inverse covariance matrix $\,(\text{\bf Cov})^{-1}\,$ encodes the degradation of the primordial signal due to the fact that, even in the absence of primordial sources, the CMB bispectrum has a variance in itself given by the six-point function of the observed CMB (see Eq.~7 of Ref.~\cite{babich:2004a}). Therefore, in the limit of weakly non-Gaussian CMB, the covariance matrix can be expressed using Wick's theorem as products of three power spectra.
The Fisher matrix element is thus computed as \footnote{Here we use the formula reported in Eq.~17 of Yadav et al.~(2008) \cite{yadav:2008a}. For the complete derivation of this formula, see appendix E of Lewis et al.~(2011) \cite{lewis:2011a}.}
\begin{align}
  \label{eq:2d_scalar_product}
  B^{(i)} \cdot B^{(j)} =
    \sum\limits_{ABC,XYZ}\;\sum\limits_{2\leq\ell_1\leq \ell_2\leq \ell_3}^{\ell_{\rm max}}
    \frac{1}{\Delta_\ldep}\;
    B^{(i),ABC}_\ldep \;
    \left(C^{-1}\right)^{AX}_\lone\;
    \left(C^{-1}\right)^{BY}_\ltwo\;
    \left(C^{-1}\right)^{CZ}_\ltre\;
    B^{(j),XYZ}_\ldep \;,
\end{align}
where the angular power spectrum $C_\ell^{XY}$ is defined as
\begin{align}
  \avg{a^X_{\ell m}\,a^{Y*}_{\ell' m'}} \;=\;
  C^{XY}_\ell\,\delta_{\ell\ell'}\,\delta_{mm'} \;.
\end{align}
$\Delta_{\ell_1 \ell_2 \ell_3} = 1,2,6$ for triangles with no, two or three equal sides and $\ell_{\rm max}$ is the maximum angular resolution attainable with the considered CMB survey.
When considering only the TTT or EEE bispectrum, the above formula reduces to the classical result of Komatsu \& Spergel (2001) \cite{Komatsu:2001a}:
\begin{align}
  B^{(i)} \cdot B^{(j)} = \sum\limits_{2\leq\ell_1\leq \ell_2\leq \ell_3}^{\ell_{\rm max}}
  \frac{1}{\Delta_\ldep}\;
  \frac{B^{(i)}_\ldep\,B^{(j)}_\ldep}{C_{\lone}\,C_{\ltwo}\,C_{\ltre}} \;.
\end{align}
Note that in our analysis we include the noise and beam contribution of the CMB survey as a linear term in the $C^{XY}_\ell$ \cite{Pogosian:2005a, knox:1995a}, and assume no correlation in noise between the temperature and polarisation detectors.
Furthermore, we shall always assume an experiment with a full sky coverage.

Using the 2D scalar product, we obtain the relation between $(\fnlgrad, \fnltime)$ and  $(\fnleq, \fnlorth)$ as \cite{Senatore:2009gt}
\begin{eqnarray}
\left(  \begin{array}{c}
  \fnleq   \\
  \fnlorth
\end{array} \right)
=
\left( \begin{array}{cc}
  \left( \frac{B^{(\rm grad)} \cdot B^{(\rm eq)}}{B^{(\rm eq)} \cdot B^{(\rm eq)}} \right) &
     \left( \frac{B^{(\rm time)} \cdot B^{(\rm eq)}}{B^{(\rm eq)} \cdot B^{(\rm eq)}} \right) \\
  \left( \frac{B^{(\rm grad)} \cdot B^{(\rm orth)}}{B^{(\rm orth)} \cdot B^{(\rm orth)}} \right) &
     \left( \frac{B^{(\rm time)} \cdot B^{(\rm orth)}}{B^{(\rm orth)} \cdot B^{(\rm orth)}} \right)
\end{array} \right)_{f_{\rm NL}=1}
\left(\begin{array}{c} \fnlgrad \\
 \fnltime
\end{array} \right).
\label{overlap}
\end{eqnarray}
We should stress that these relations are obtained assuming that only one type of the templates is present at the same time, that is, we have either $\fnleq$ or $\fnlorth$.
This is the same definition used by the WMAP team when they quote constraints on these non-linear parameters.

In this paper, we are interested in a joint analysis where both equilateral and orthogonal non-Gaussianity exist. We therefore introduce a new set of non-linear parameters, $\hat{f}^i= (\hfnleq, \hfnlorth)$, which are related to $f_i=(\fnleq, \fnlorth)$ as
\begin{equation}
\sum_j F_{ij} \hat{f}^j = F_{ii} f_i, \quad i={\rm eq, orth},
\label{relation}
\end{equation}
where the template Fisher matrix $F_{ij}$,
\begin{eqnarray}
F &=& \left( \begin{array}{cc}
 B^{\rm (eq)} \cdot B^{\rm (eq)} & B^{\rm (eq)} \cdot B^{\rm (orth)} \\
 B^{\rm (eq)} \cdot B^{\rm (orth)} & B^{\rm (orth)} \cdot B^{\rm (orth)}
\end{array} \right)_{f_{\rm NL}=1} \;,
\label{eq:template_fisher}
\end{eqnarray}
encodes the overlap between the two observational templates in $\ell$-space. The new parameters, $\hat{f}^i$, take into account the contamination from the other type of non-Gaussianity; they are equivalent to $f_i$ only if there is no correlation between the two estimators, that is if $\,r= F_{ij}/(F_{ii} F_{jj})^{1/2}=0\,$.

Using Eqs.~(\ref{overlap}), (\ref{relation}) and (\ref{eq:template_fisher}), we obtain the relation between the model parameters $(\fnlgrad, \fnltime)$ and $(\hfnleq, \hfnlorth)$ as \cite{Senatore:2009gt}
\begin{eqnarray}
\left(  \begin{array}{c}
  \hfnleq   \\
  \hfnlorth
\end{array} \right)
= F^{-1} M
\left(\begin{array}{c} \fnlgrad \\
 \fnltime
\end{array} \right), \quad
M= \left( \begin{array}{cc}
   B^{(\rm grad)} \cdot B^{(\rm eq)} & B^{(\rm time)} \cdot B^{(\rm eq)} \\
   B^{(\rm grad)} \cdot B^{(\rm orth)} & B^{(\rm time)} \cdot B^{(\rm orth)}
\end{array} \right)_{f_{\rm NL}=1} \;,
\label{overlap2}
\end{eqnarray}
where $M$ is the overlap matrix between the theoretical shapes and the observational templates.
Given the best-fit values of $\hat{f}^i$ from the data, $\hat{f}^i_{\rm best}$, and the associated covariance matrix $C=F^{-1}$, we define a $\chi^2$ statistic for model parameters $\hat{f}^i$ as
\begin{equation}
\chi^2 \,=\, (\hat{f}^{i} - \hat{f}^{i}_{\rm best}) \, C^{-1}_{ij} \, (\hat{f}^{j} - \hat{f}^{j}_{\rm best}) \,=\, (\hat{f}^{i} - \hat{f}^{i}_{\rm best}) \, F_{ij} \, (\hat{f}^{j} - \hat{f}^{j}_{\rm best}).
\end{equation}
This $\chi^2$ statistic quantifies the agreement between the observed bispectrum and the model bispectrum.

\section{Constraints from Planck}

In this section, we use Planck's measurement of $(\hfnleq, \hfnlorth)$ to constrain the theoretical parameters in the DBI galileon model. The Planck collaboration gave constraints on the equilateral and orthogonal non-Gaussianity at the 1$\sigma$ level as \cite{planck-ng}
\begin{equation}
\fnleq= -42 \pm 75, \quad \fnlorth= -25 \pm 39.
\label{Planckconstraints}
\end{equation}
Note that these constraints do not include polarisation yet. We discuss how polarisation will improve the constraints later in this section.

We utilise the Fisher module of the Second Order Non-Gaussianity (\textsf{SONG}) code \cite{Pettinari:2013he} to numerically obtain the $F$ and $M$ matrices which are needed to relate $(\hfnleq, \hfnlorth)$ to $(\fnlgrad, \fnltime)$ through Eq.~(\ref{overlap2}). In doing so, we take into account the expected sensitivity and noise properties of Planck\footnote{We plan to release the code that we used for this analysis as a separate module for the Boltzmann code CLASS \cite{blas:2011a} in 2014 \cite{Pettinari:2013a}.}.

In order to obtain the Fisher matrices, we first estimate the bispectrum integral in Eq.~\eqref{eq:bispectrum_integral} for the four shapes considered in this paper: equilateral, orthogonal, and the two galileon shapes in Eq.~\eqref{eq:shapeone}. The equilateral and orthogonal templates are separable in $(k_1,k_2,k_3)$, meaning that their computation can be conveniently split into one-dimensional integrations. The galileon shapes do not have this desirable property and we treat them as described in Sec.~5 of Ref.~\cite{Pettinari:2013he}. We obtain the temperature transfer functions $\Delta_{l}(k)$ with CLASS \cite{lesgourgues:2011a, blas:2011a} by employing a LCDM model with Planck cosmological parameters (Planck+WP+highL+BAO) \cite{planck-parameters} whereby $h = 0.677$, $\Omega_b = 0.0483$, $\Omega_{\text{cdm}} = 0.259$, $\Omega_\Lambda = 0.693$, $A_s = 2.214\times10^{-9}$, $\tau_\text{reio} = 0.0952$, $N_\text{eff} = 3.04$.
Note that, since the galileon shapes were computed assuming slow-roll conditions, we consistently set $n_s = 1$ also for the equilateral and orthogonal templates. We have checked that this assumption does not affect our conclusions\footnote{Repeating our analysis using the measured value from Planck ($n_s=0.96$) yields the same parameter constraints within $\sim 2\%$ accuracy.
This small difference is partly explained by the fact that the constraints are obtained using ratios of Fisher matrix elements, so that the effect of small variations in the cosmological parameters tends to cancel.}.

We compute the full $4\times 4$ Fisher matrix in Eq.~\eqref{eq:2d_scalar_product} by interpolating our four numerical bispectra on a mesh in ($\ell_1,\ell_2,\ell_3$) \cite{Pettinari:2013a,Pettinari:2013he}. We employ the noise model described in Ref.~\cite{Pogosian:2005a} and consider the $100, 143, 217\,\,\text{GHz}$ frequency channels of the Planck experiment, with noise and beam parameters from Ref.~\cite{planck-hfi-core-team:2013a} and $\lmax=2000$. When considering only the temperature bispectrum TTT, we thus obtain the following full Fisher matrix:
\begin{equation}
  \label{eq:full_fisher_planck}
F^T_{\,\text{full}} \,=\, \left( \begin{array}{rrrr}
    2.38   &  -0.208  &    2.51  &   3.06 \\
   -0.208  &  8.47    &  -0.708  &  -2.49 \\
     2.51  &  -0.708  &    2.68  & 3.39 \\
   3.06   &  -2.49  &    3.39  & 4.67
\end{array} \right) \times 10^{-4} \;,
\end{equation}
where the ordering of the rows and columns is $\fnleq$, $\fnlorth$, $\fnlgrad$, $\fnltime$. The $F_{\,\text{full}}$ matrix contains all the information needed to produce Planck constraints on the parameters of the DBI galileon model. It also encodes the correlations in $\ell$-space between the two considered galileon shapes (lower-right submatrix).

The $F$ matrix is the upper-left submatrix of the full Fisher matrix
\begin{equation}
F = \left( \begin{array}{cc}
 B^{\rm (eq)} \cdot B^{\rm (eq)} & B^{\rm (eq)} \cdot B^{\rm (orth)} \\
 B^{\rm (eq)} \cdot B^{\rm (orth)} & B^{\rm (orth)} \cdot B^{\rm (orth)}
\end{array} \right)_{f_{\rm NL}=1}
=
\left( \begin{array}{cc}
 2.38 & -0.208 \\
 -0.208 & 8.47
\end{array} \right) \times 10^{-4} \;.
\end{equation}
The errors on $\fnleq$ and $\fnlorth$ can be obtained as $1/\sqrt{F_{ii}}$, giving
$\,\Delta \fnleq = 64.8\,$ and $\,\Delta \fnlorth = 34.4\,$.
These errors are roughly $15\%$ smaller than the actual constraints obtained by the Planck collaboration \cite{planck-ng}.
The reason is that the error budget in Planck's analysis includes uncertainties from more subtle effects such as incomplete foreground removal. Furthermore, we are using $n_s=1$ rather than the measured value of $n_s=0.96$, for reasons explained above.
The bispectrum Fisher matrix was not provided by the Planck collaboration, thus we will use our Fisher matrix to constrain the parameters.

\begin{figure}[ht]
  \centering{
  \includegraphics[width=6in]{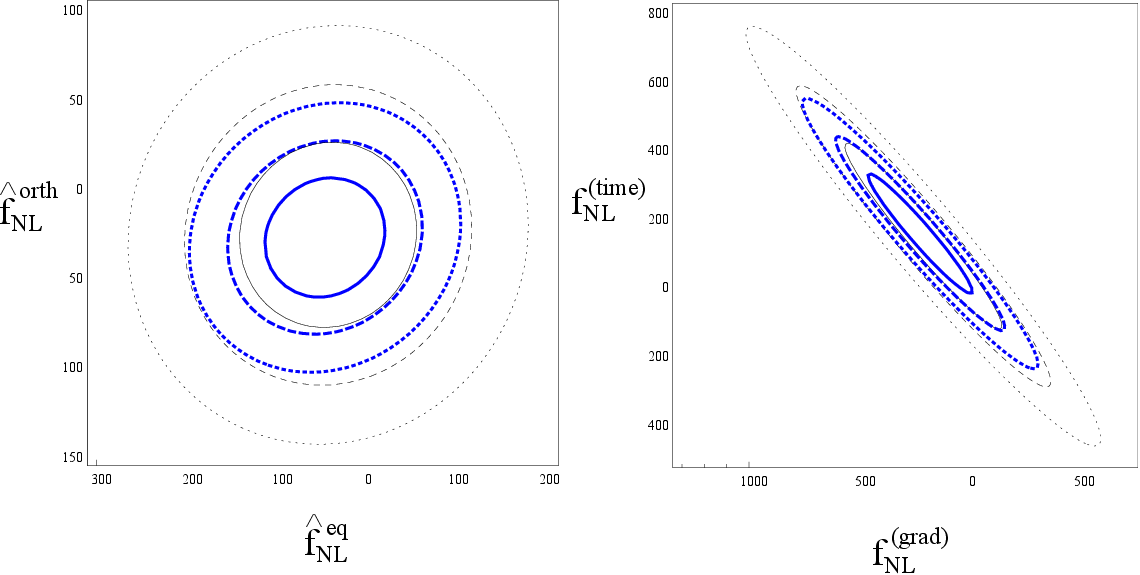}
  }
  \caption{Planck constraints on $(\fnleq, \fnlorth)$ in the left panel and on $(\fnlgrad,\fnltime)$ in the right panel. $68\%$ (black, solid), $95\%$ (black, dashed) and $99.7\%$ (black, dotted) confidence contours are shown. The forecasts for constraints including polarisation data are shown in blue.
  }
  \label{fig:planck_constraints_fnl}
\end{figure}

\begin{figure}[ht]
  \centering{
  \includegraphics[width=6in]{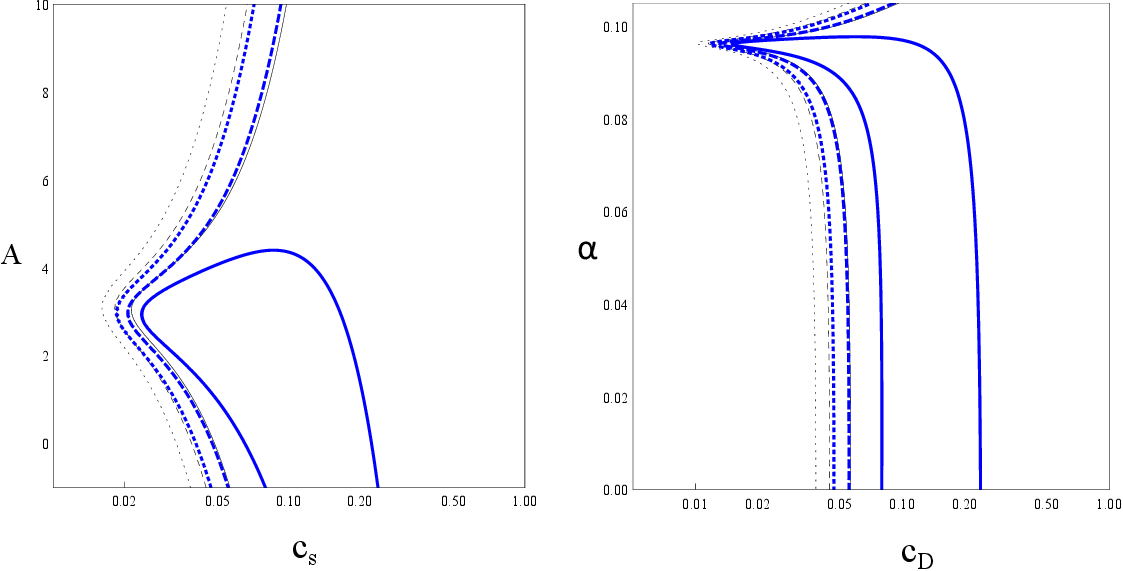}
  }
  \caption{Planck constraints on $(c_s, A)$ in the left panel and on $(\alpha, c_D)$ in the right panel. Confidence contours are the same as in Fig.~1. The forecasts for constraints including polarisation data are shown in blue.}
  \label{fig:planck_constraints_parameters}
\end{figure}

The $M$ matrix, that is the overlap between theoretical shapes and templates, is the upper-right submatrix of the full Fisher matrix in Eq.~\eqref{eq:full_fisher_planck}:
\begin{equation}
M= \left( \begin{array}{cc}
   B^{(\rm grad)} \cdot B^{(\rm eq)} & B^{(\rm time)} \cdot B^{(\rm eq)} \\
   B^{(\rm grad)} \cdot B^{(\rm orth)} & B^{(\rm time)} \cdot B^{(\rm orth)}
\end{array} \right)_{f_{\rm NL}=1}
=
\left( \begin{array}{cc}
 2.51 & 3.06 \\
 -0.708 & -2.49
\end{array} \right) \times 10^{-4} \;.
\end{equation}
Then the relation between $(\fnlgrad, \fnltime)$ and  $(\hfnleq, \hfnlorth)$ is given by Eq.~(\ref{overlap2}) as
\begin{eqnarray}
\left(  \begin{array}{c}
  \hfnleq   \\
  \hfnlorth
\end{array} \right)
= \left( \begin{array}{cc}
 1.050 & 1.263 \\
 -0.0578 & -0.263
\end{array} \right)
\left(\begin{array}{c} \fnlgrad \\
 \fnltime
\end{array} \right) .
\label{FisherPlanck}
\end{eqnarray}
The Planck collaboration provided a relation between the theoretical parameters and $(\hfnleq, \hfnlorth)$ in terms of the parameters of the effective field theory ($c_s, A$) \cite{planck-ng}:
\begin{eqnarray}
  \nonumber
&\hfnleq =& \frac{1-c_s^2}{c_s^2}(-0.275 + 0.0780 A), \\
&\hfnlorth =& \frac{1-c_s^2}{c_s^2}(0.0159 -0.0167 A),
\label{effectivePlanck}
\end{eqnarray}
which gives
\begin{eqnarray}
\left(  \begin{array}{c}
  \hfnleq   \\
  \hfnlorth
\end{array} \right)
= \left( \begin{array}{cc}
 1.048 & 1.264 \\
 -0.0606 & -0.271
\end{array} \right)
\left(\begin{array}{c} \fnlgrad \\
 \fnltime
\end{array} \right) .
\label{FisherPlanck2}
\end{eqnarray}
This is consistent with Eq.~(\ref{FisherPlanck}) within a few percentage level.

We define a $\chi^2$ statistic
\begin{equation}
\chi^2 = (\hat{f}^{i} - \hat{f}^{i}_{\rm Planck}) F_{ij} (\hat{f}^{j} - \hat{f}^{j}_{\rm Planck}).
\end{equation}
The central values $\hat{f}^i_{\rm Planck}$ are obtained by substituting the values of Eq.~(\ref{Planckconstraints}) into Eq.~(\ref{relation}), and by making use of the template Fisher matrix in Eq.~(\ref{FisherPlanck}):
\begin{equation}
\sum_j F_{ij} \hat{f}^j_{\rm  Planck} = F_{ii} f_{i \; \rm  Planck}, \quad f_{i \; \rm Planck}
=(-42, -25) \;.
\end{equation}
This yields the following central values for the non-linear parameters
\begin{equation}
(\hfnleq, \hfnlorth)
= (-44, \,\,-26)\,, \quad (\fnlgrad, \fnltime)
=(-219, \,\,147) \;.
\label{Planckcentral}
\end{equation}

In Fig.~\ref{fig:planck_constraints_fnl}, the left panel shows 68$\%$, 95$\%$ and 99.7$\%$ confidence regions in the $(\hfnleq, \hfnlorth)$ plane defined by threshold $\chi^2$ values 2.28, 5.99 and 11.62. The right panel shows the same confidence regions in the  $(\fnlgrad, \fnltime)$ plane.
In Fig~\ref{fig:planck_constraints_parameters}, these constraints are shown in the $(c_s, A)$ plane in the effective theory and the $(c_D, \alpha)$ plane in the DBI galileon model. At the 95$\%$ confidence level, we only obtain the lower bound for the sound speed $c_s > 0.02$ or $c_D >0.01$.

We now discuss how much the constraint will be improved by the addition of Planck's polarisation maps, which are expected to be released by the end of 2014. We combine the eight temperature and polarisation bispectra (TTT, TTE, TET, ETT, EET, ETE, TEE, EEE) using the procedure outlined in section \ref{sec:cmb_bispectra}.
For the E-mode noise sensitivity, we assume a variance four times larger than that in the temperature\footnote{The E-mode polarisation is obtained as a linear combination of two measured Stokes parameters ($Q$ and $U$), while the temperature is simply proportional to the measured intensity ($I$), thus providing a factor two degradation in the E-modes noise variance. Furthermore, Planck has half the number of detectors in polarisation than in temperature, hence another factor 2.}, and thus obtain the following full Fisher matrix:
\begin{equation}
  \label{eq:full_fisher_planck_polarisation}
F^{T+E}_{\,\text{full}} \,=\, \left( \begin{array}{rrrr}
    5.22   &  -1.09  &    5.56  &   6.97 \\
   -1.09  &  20.6    &  -2.37  &  -6.90 \\
   5.56  &  -2.37  &    6.02  & 7.86 \\
   6.97   &  -6.90  &   7.86  & 11.1
\end{array} \right) \times 10^{-4} \;.
\end{equation}
From the diagonal elements in the first two lines, it follows that the errors on $\fnleq$ and $\fnlorth$ will be improved to $\,\Delta \fnleq = 44\,$ and $\,\Delta \fnlorth = 22\,$ at the $1\sigma$ level.
Assuming that the central values of these two parameters will not change with the addition of polarised data, we can use the above Fisher matrix to estimate how much Planck polarisation will improve the forecasts on the galileon model.
The corresponding contour lines are shown in Figures~\ref{fig:planck_constraints_fnl} and~\ref{fig:planck_constraints_parameters} in blue; the latter suggests that Planck polarisation measurements may provide a hint for a non-canonical sound speed at the 68$\%$ confidence level.

\begin{figure}[ht]
  \centering{
  \includegraphics[width=6in]{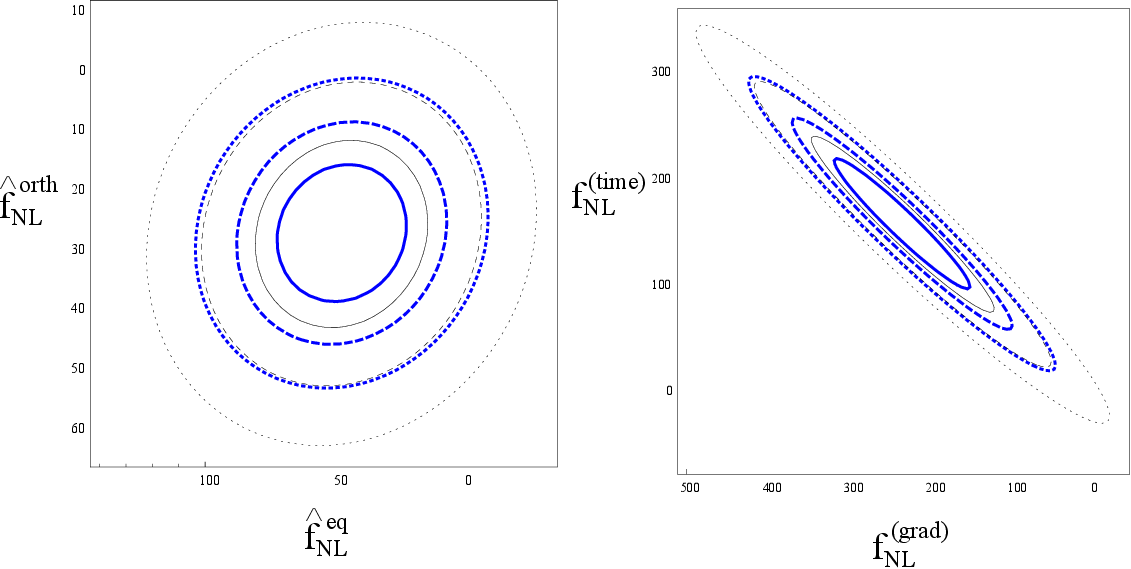}
  }
  \caption{Forecasts for constraints on $(\fnleq, \fnlorth)$ in the left panel and $(\fnlgrad,\fnltime)$ in the right panel. $68\%$ (solid), $95\%$ (dashed) and $99.7\%$ (dotted) confidence contours are shown for COrE (black, thin) and PRISM (blue, thick). }
  \label{fig:future_constraints_fnl}
\end{figure}

\begin{figure}[ht]
  \centering{
  \includegraphics[width=6in]{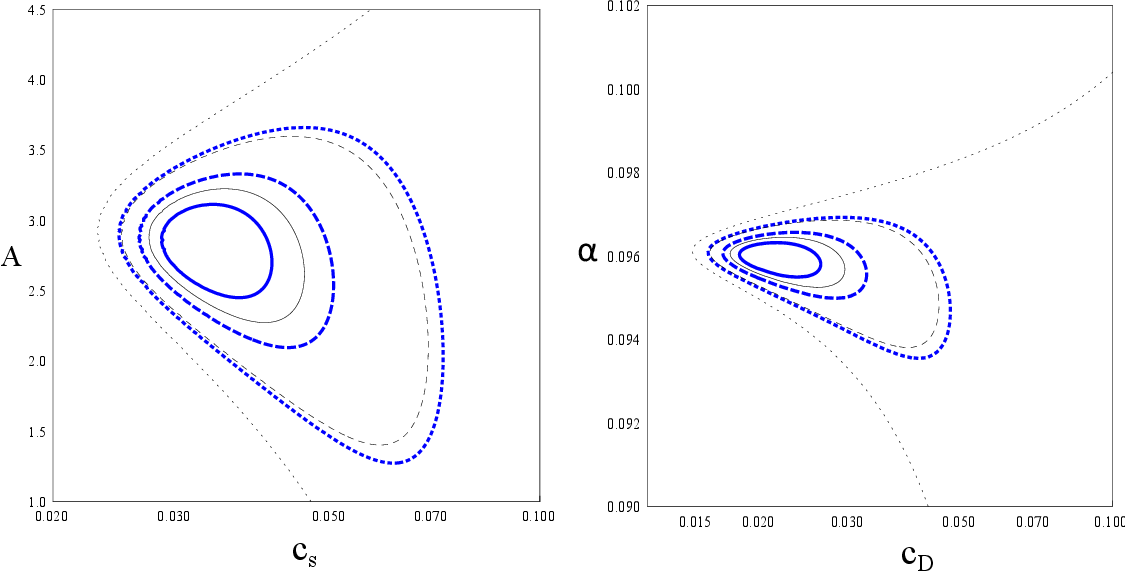}
  }
  \caption{Forecasts for constraints on $(c_s, A)$ in the left panel and $(\alpha, c_D)$ in the right panel. Confidence contours are the same as in Fig.~3.}
  \label{fig:future_constraints_parameters}
\end{figure}

\section{Forecasts for COrE and PRISM}
\label{sec:core_prism_forecasts}
The results obtained in the previous section suggest that further improvements on measurements of $\,(\hfnleq, \hfnlorth)\,$ can strongly constrain DBI galileon models.
In this section, we estimate how well post-Planck measurements on $(\hfnleq, \hfnlorth)$ will constrain the theoretical parameters in the DBI galileon model.

We consider two satellite CMB experiment that were recently proposed: the Cosmic Origins Explorer (COrE) \cite{core-collaboration:2011} and the Polarized Radiation Imaging and Spectroscopy Mission (PRISM) \cite{prism-collaboration:2013b}. Both experiments represent a substantial improvement over Planck as they will feature more frequency channels and detectors, together with an increased angular resolution.
For CoRE, we consider the $105, 135, 165, \allowbreak195, 225, 255, 285\,\,\text{GHz}$ frequency channels with the noise and beam values reported in Ref.~\cite{core-collaboration:2011} and $\lmax=3000$. After running \textsf{SONG} with these parameters, we find the following full Fisher matrix:
\begin{equation}
  \label{eq:full_fisher_COrE}
F^{T+E}_{\,\text{full}} \,=\, \left( \begin{array}{rrrr}
    2.15  & -0.49  &   2.30  &  2.90 \\
   -0.49  &  9.35  &  -1.06  & -3.10 \\
    2.30  & -1.06  &   2.49  &  3.28 \\
    2.90  & -3.10  &   3.28  &  4.72
\end{array} \right) \times 10^{-3} \;.
\end{equation}
For PRISM we consider the $105, 135, 160, \allowbreak185, 200, 220, 265\,\,\text{GHz}$ channels with the noise and beam values in Ref.~\cite{prism-collaboration:2013b} and $\lmax=3000$, thus obtaining:
\begin{equation}
  \label{eq:full_fisher_PRISM}
F^{T+E}_{\,\text{full}} \,=\, \left( \begin{array}{rrrr}
   3.81  &  -0.86  &   4.07  &   5.13 \\
  -0.86  &   17.4  &  -1.90  &  -5.65 \\
   4.07  &  -1.90  &   4.43  &   5.81 \\
   5.13  &  -5.65  &   5.81  &   8.39
\end{array} \right) \times 10^{-3} \;.
\end{equation}
It follows that the errors on $\fnleq$ and $\fnlorth$ will be improved as $\,\Delta \fnleq = 21.6\,$ and $\,\Delta \fnlorth = 10.3\,$ for COrE, and $\,\Delta \fnleq = 16.2\,$ and $\,\Delta \fnlorth = 7.6\,$ for PRISM at the $1\sigma$ level. These numbers should be compared with the values that we have obtained for Planck: $\,\Delta\fnleq = 44\,$ and $\,\Delta \fnlorth = 22\,$.
Using Eq.~(\ref{overlap2}), one can therefore find the relation between $(\fnlgrad, \fnltime)$ and $(\hfnleq, \hfnlorth)$:
\begin{eqnarray}
\left(  \begin{array}{c}
  \hfnleq   \\
  \hfnlorth
\end{array} \right)
= \left( \begin{array}{cc}
 1.057 & 1.289 \\
 -0.058 & -0.264
\end{array} \right)
\left(\begin{array}{c} \fnlgrad \\
 \fnltime
\end{array} \right),
\label{FisherPlanckpol}
\end{eqnarray}
for COrE and
\begin{eqnarray}
\left(  \begin{array}{c}
  \hfnleq   \\
  \hfnlorth
\end{array} \right)
= \left( \begin{array}{cc}
 1.055 & 1.288 \\
 -0.057 & -0.261
\end{array} \right)
\left(\begin{array}{c} \fnlgrad \\
 \fnltime
\end{array} \right),
\label{FisherCoRE}
\end{eqnarray}
for PRISM. Although these numerical coefficients are specific to experiments, they are all very similar. This is partly explained by the fact that the coefficients are obtained as ratios of Fisher matrix elements, \ie they measure an overlap rather than an amplitude.

In Fig.~\ref{fig:future_constraints_fnl}, the left panel shows 68$\%$, 95$\%$ and 99.7$\%$ confidence regions in the
$(\hfnleq, \hfnlorth)$ plane defined by threshold $\chi^2$ values 2.28, 5.99 and 11.62 for COrE (thin, black) and PRISM (thick, blue). The right panel shows the same confidence regions in the  $(\fnlgrad, \fnltime)$ plane.
In Fig.~\ref{fig:future_constraints_parameters}, the same constraints are shown in the $(c_s, A)$ plane in the effective theory and in the $(c_D, \alpha)$ plane in the DBI galioen model.
We find that COrE and PRISM will not only confirm a non-canonical sound speed but also exclude the conventional DBI inflation model at more than the 95$\%$ and 99$\%$ confidence level respectively, assuming that the central values found in Planck will not change.

\section{Conclusion}
In this paper, we obtained constraints on the two parameters $(\alpha, c_D)$ in the single-field DBI galioen model using Planck results. The parameter $\alpha$ parametrises the effect of the induced gravity on a
brane and describes the deviation from the conventional DBI inflation model, while
$c_D$ becomes the sound speed in the DBI inflation limit. The bispectrum of the Newtonian potential in this model is not separable and it is therefore numerically challenging to construct an optimal estimator. Therefore, we used the \textsf{SONG} code \cite{Pettinari:2013he, Fidler:2014a} to obtain the relation between the amplitudes of theoretical bispectra $(\fnlgrad, \fnltime)$ and the equilateral and orthogonal observational templates $(\fnleq,\fnlorth)$ by properly taking into account the sensitivity and noise properties of Planck. We then used the bispectrum Fisher matrix for $(\fnleq,\fnlorth)$ to obtain the constraints on the two parameters. Using the central values for equilateral and orthogonal non-Gaussianities found in the temperature data from the Planck survey \cite{planck-ng}, we obtained the lower bound for $c_D$ as $c_D > 0.01$ at the 95$\%$ confidence level.

We also included polarisation in the bispectrum Fisher matrix and provided forecasts for the upcoming Planck polarised data and for two proposed post-Planck experiments, CoRE and PRISM, by properly taking into account the noise sensitivity and resolution properties of these experiments. By assuming that the central values found in Planck temperature data remain the same, we found that Planck polarisation measurements may provide a hint for a non-canonical sound speed at the 68$\%$ confidence level. COrE and PRISM will not only confirm a non-canonical sound speed but also exclude the conventional DBI inflation model at more than the 95$\%$ and 99$\%$ confidence level, respectively. This indicates that improving constraints on non-Gaussianity further by future CMB experiments is still invaluable to constrain physics of the early universe.

\subsection*{Acknowledgments}
We would like to thank S. Renaux-Petel for useful discussions. KK thanks G-B. Zhao for discussions on the Fisher matrix. KK and GWP were supported by the Leverhulme trust. GWP also acknowledges support from the UK Science and Technology Facilities Council grant number ST/I000976/1. KK and CF are supported by the UK Science and Technology Facilities Council grants number ST/K00090/1 and ST/L005573/1. SM is supported by CNRS, France.

\bigskip

\end{document}